\documentclass[aps,prb, preprint, citeautoscript, showpacs,floatfix,superscriptaddress]{revtex4-1}
\usepackage{bm}
\usepackage{amssymb}
\usepackage{graphicx}
\usepackage{amsmath}
\usepackage{textcomp}
\usepackage{multirow}

\begin{document}

\title{Third order optical nonlinearity of three dimensional massless Dirac fermions}
\author{J. L. Cheng}
\email{jlcheng@ciomp.ac.cn}
\affiliation{The Guo China-US Photonics Laboratory, State Key Laboratory of Applied Optics, Changchun Institute of Optics, Fine Mechanics and Physics, Chinese Academy of  Sciences, Changchun 130033, China.}
\affiliation{School of Physical Sciences, University of Chinese Academy of Sciences, Beijing 100049, China}
\author{J. E. Sipe}
\affiliation{Department of Physics, University of Toronto, Toronto, Ontario, Canada}
\author{S. W. Wu}
\affiliation{State Key Laboratory of Surface Physics, Key Laboratory
  of Micro and Nano Photonic Structures (MOE), and Department of
  Physics, Fudan University, Shanghai 200433, China}

\begin{abstract}
We present analytic expressions for the electronic
contributions to the linear conductivity $\sigma^{(1)}_{3d}(\omega)$ and
the third order optical conductivity
$\sigma^{(3)}_{3d}(\omega_1,\omega_2,\omega_3)$  of  
three dimensional massless Dirac fermions, the
quasi-particles  relevant for the low
energy excitation of topological Dirac semimetals and Weyl semimetals. Although
there is no gap for massless Dirac fermions, a finite chemical potential $\mu$
can lead to an effective gap parameter, which plays an important role in
the qualitative features of interband optical transitions. For gapless linear
dispersion in three dimension, the imaginary part of the linear conductivity diverges as a logarithmic function of the cutoff energy, while the real
part is linear with  photon frequency $\omega$ as
$\hbar\omega>2|\mu|$. The third order conductivity exhibits features very
similar to those of two dimensional Dirac fermions, {\it
  i.e.}, graphene,  but with the amplitude for a single Dirac cone generally two orders of
magnitude smaller in three dimension than in two dimension. There are many resonances associated with
the chemical potential induced gap parameters, and divergences
associated with the intraband transitions. The details of the third
order conductivity are discussed for third harmonic generation, the Kerr effect and two-photon
carrier injection, parametric frequency conversion, and two-color
coherent current injection. Although the expressions we derive are
limited to the clean limit at zero temperature, the generalization
to include phenomenological relaxation processes at finite temperature is
straightforward and is presented. 
\end{abstract}
 
\maketitle
\section{Introduction}
Two dimensional (2D) massless Dirac fermions (DFs) have been
investigated extensively in condensed matter systems since their first experimental
realization in graphene, and their properties are significantly
different than those of fermions in the more usual parabolic bands
\cite{Rev.Mod.Phys._81_109_2009_CastroNeto,Nanoscale_7_4598_2015_Ferrari}. Their
attractive 
optical properties \cite{Nat.Photon._4_611_2010_Bonaccorso} include  broadband linear optical 
absorption and the ability to use the chemical potential to tune both
plasmon resonances and an extremely strong nonlinear optical response \cite{Phys.Rep._535_101_2014_Glazov}.
The strong nonlinear response makes graphene a potential candidate for
integration in photonic devices
\cite{Proc.R.Soc.A_473_20170433_2017_Ooi,Nat.Photon._6_554_2012_Gu,Nat.Commun._9_2675_2018_Vermeulen}
as a source of nonlinear functionality, and it has been the focus of a large number of
experimental \cite{Nat.Photon.___2018_Jiang,Nat.Nano._X_X_2018_Soavi} and
theoretical
\cite{Europhys.Lett._79_27002_2007_Mikhailov,NewJ.Phys._16_53014_2014_Cheng,*Corrigendum_NewJ.Phys._18_29501_2016_Cheng,Phys.Rev.B_91_235320_2015_Cheng,*Phys.Rev.B_93_39904_2016_Cheng,Phys.Rev.B_92_235307_2015_Cheng,Phys.Rev.B_93_085403_2016_Mikhailov,Phys.Rev.B_93_161411_2016_Rostami,Phys.Lett.A_380_304_2016_Margulis,Phys.Rev.B_99_195407_2019_Hipolito,2DMater._6_031003_2019_Semnani}
studies over the past decade.
Experiments have explored different nonlinear phenomena including
third harmonic generation (THG), the Kerr effect
and two photon carrier injection, parametric frequency conversion (PFC), and
two-color coherent current injection (CCI); the corresponding nonlinear coefficients
have been extracted for different photon energies and chemical
potentials. Theoretical studies have been mainly at the level of
independent particle approximation, and have presented perturbative
expressions and numerical simulations. Recently, 
many-body effects \cite{Phys.Rev.B_95_35416_2017_Rostami,Phys.Rev.B_97_115454_2018_Avetissian,Phys.Rev.B_100_245433_2019_Cheng}
have been shown to play a significant role in the nonlinear optical response.
And in the development of theories of topological materials, 2D
massless DFs have been shown to determine the properties of the low energy excitation of surface
states of a topological insulators, despite the small energy range
over which the linear dispersion approximation is valid. 

In a two band model for 2D DFs, a mass can
be introduced. The resulting dispersion relation can be realized around the
band edge of gapped graphene, or around the band edge of a monolayer of BN
or MoS$_2$, and in other 2D materials. The optical
nonlinearities of 2D massive DFs have also
been investigated both experimentally and theoretically. Jafari
\cite{J.Phys.Condens.Matter._24_205802_2012_Jafari} presented a theory
for THG using a Feynman diagrammatic technique, describing the light-matter interaction in the framework of a vector potential. However,
in the limit of vanishing mass his result does not converge to the
results of other
studies\cite{NewJ.Phys._16_53014_2014_Cheng,*Corrigendum_NewJ.Phys._18_29501_2016_Cheng}. Cheng 
{\it et al.} investigated various nonlinear effects both by 
numerically solving the equations of motion
\cite{Phys.Rev.B_92_235307_2015_Cheng} and by approximation from the
results of gapped graphene under a 
perpendicular magnetic field \cite{Phys.Rev.B_97_125417_2018_Cheng}. Recently, we derived  analytic
expressions for the third order conductivities of gapped
graphene \cite{APLPhotonics_4_034201_2019_Cheng} at general
frequencies, following earlier work on graphene \cite{Phys.Rev.B_91_235320_2015_Cheng,*Phys.Rev.B_93_39904_2016_Cheng}. 

There have also been a host of recent studies focused on the
prediction and discovery of {\it three} dimensional (3D) Dirac and Weyl
semimetals \cite{Phys.Rev.Lett._108_140405_2012_Young,2DMaterials_6_32001_2019_Ma,Annu.Rev.Condens.MatterPhys_9_359_2018_Burkov,Annu.Rev.Mater.Res._49_153_2019_Gao,Nat.Mater._15_1140_2016_Jia,arXiv_1603.02821,Annu.Rev.Condens.MatterPhys_8_337_2017_Yan,SPIN_06_1640003_2016_Yang},
where the low energy excitations can be described by DFs  with a {\it three} dimensional wave vector. As an analogue of
2D massless DFs, 3D massless DFs \cite{SPIN_06_1640003_2016_Yang} possess gapless linear
dispersion and an interesting band
topology around the Dirac point, which leads to extraordinary optical
properties. As well, the chiral anomaly in Weyl semimetals can be probed with the presence of both the electric
field and magnetic field \cite{Natl.Sci.Rev._6_206_2018_Moore}. The nonlinear optical properties of
3D massless DFs have also attracted attention \cite{Nat.Phys._13_842_2017_Ma,Nature_565_337_2018_Ma,Nature_565_337_2018_Ma,Ann.Phys._529_1600359_2017_Chi,Nat.Commun._8_15995_2017_Juan,Phys.Rev.Lett._115_216806_2015_Sodemann,Phys.Rev.Lett._111_027201_2013_Vazifeh,arXiv_1712.09363v2}.
Experimentally, huge nonlinear optical
coefficients\cite{Nat.Mater._18_476_2019_Ma} have been observed, although probably at 
frequencies much higher than those at which the linear
dispersion approximation is valid. There are interesting recent
theoretical
predictions\cite{APLPhotonics_4_034402_2019_Ooi,Opt.Express_27_38270_2019_Zhang,Phys.B_555_81_2019_Zhong}
for the Kerr effect and THG, both within the 
framework of the Boltzmann equation and in a treatment including intraband and interband transitions. In these studies the
focus was on frequencies in the terahertz regime, and
possible applications in terahertz plasmonics have been investigated
\cite{Opt.Commun._462_125319_2020_Ooi}. However, the light-matter
interaction was described in a velocity gauge, and additional care may be required
to confirm that no unphysical divergences have been induced by  band
truncation; a treatment based on the length gauge \cite{Phys.Rev.B_48_11705_1993_Sipe,Phys.Rev.B_96_195413_2017_Taghizadeh}, where such
difficulties are not present, is clearly in order. Further, in order to extend the application of these
materials to various nonlinear optical scenarios, it would be helpful
to understand the general frequency dependence of the third order
conductivity, especially in a comparison with that of graphene; this
has not yet been done.

In this work, we derive analytic expressions for linear and third order
optical conductivities of 3D massless DFs. Our strategy is based on
employing earlier results found for the linear and nonlinear optical
response of gapped graphene. In fact, we show that the response
coefficients for 3D massless DFs can be written as an integral over
the results for gapped graphene with different gaps.  Our treatment includes the intraband and interband optical
transitions, in a framework where the light-matter
interaction is described in the length gauge. Our expressions for the third
order conductivities describe a
general input frequency dependence for the clean limit at zero
temperature. After analyzing the structures of the conductivities, we discuss in detail
the coefficients for THG, the Kerr effect and two
photon carrier injection, PFC, and
two-color CCI. To better understand of the
physics of the nonlinear processes, comparisons with that of graphene
are made.

We organize the paper as following: in Section~\ref{sec:connections} we summarize the symmetries of frequency dependence of the linear and nonlinear conductivities of 2D massive DFs. In Section~\ref{sec:sigmadf} we describe how to 
construct the conductivity of 3D massless DFs
from the conductivity of 2D massive DFs, and present the analytic
expressions for linear conductivity 
and third order conductivity; in
Section~\ref{sec:results} we discuss the details of the conductivities for different
optical phenomena, including the linear optical response, THG, the Kerr effect and two photon carrier injection,
PFC, and two-color CCI; in Sec.~\ref{sec:conclusion} we discuss and
conclude, indicating how the extension of our results to include
finite temperature and a phenomenological description of relaxation
processes can easily be implemented.  

\section{Conductivities for 2D Dirac Fermions\label{sec:connections}}
Two dimensional massive DFs in one Dirac cone can be described by the Hamiltonian
\begin{align}
  H_{2d}(\bm \kappa,\Delta)= \hbar v_F \bm
  \kappa\cdot\bm\sigma + \Delta \sigma_z\,.\label{eq:h2d}
\end{align}
where $v_F$ is the Fermi velocity, $\bm\sigma =\sigma_x\hat{\bm  x}+\sigma_y\hat{\bm y}+\sigma_z\hat{\bm z}$ has its components as Pauli matrices, $\bm\kappa=\kappa_x \hat{\bm x}+\kappa_y\hat{\bm y}$ is a
two-dimensional wave vector, and $\Delta$ is a mass parameter to give
a gap $2|\Delta|$ at the Dirac point. Depending on the material, there
can exist multiple Dirac cones, and for different materials the model
Hamiltonian can take in  different
forms. For example, the low energy excitations of gapped 
graphene are described by the Hamiltonian
\begin{align}
  H_{gg;\tau}( \bm \kappa, \Delta) =  \hbar v_F (\tau \kappa_y \sigma_x - \kappa_x
  \sigma_y) + \Delta \sigma_z  \,, \label{eq:gg2}
\end{align}
where $\tau=\pm$ is a valley index for two different Dirac
cones.

For such Hamiltonians, we consider the linear optical conductivity
tensor $\sigma^{(1);da}(\omega)$ and third order optical conductivity
tensor $\sigma^{(3);dabc}(\omega_1,\omega_2,\omega_3)$, where the Roman
letters $d$, $a$, $b$, $c$ refer to the Cartesian directions,  and $\omega$ and $\omega_i$ refer
to the optical frequencies.  The second order
response vanishes in the dipole approximation, as we discuss below.  The results of gapped graphene have been
given earlier \cite{APLPhotonics_4_034201_2019_Cheng}, and will be summarized in the following. 

\subsection{Symmetry properties of conductivities for two dimensional
  massive Dirac fermions}
We denote the conductivities for a 2D Dirac cone by $\sigma^{(1);da}_{2d}(\omega)$ and
$\sigma^{(3);dabc}_{2d}(\omega_1,\omega_2,\omega_3)$. The
Hamiltonian $H_{2d}(\bm\kappa,\Delta)$ satisfies the rotational
symmetry condition
\begin{align}
  U_\theta  H_{2d}(R_\theta\bm\kappa,\Delta) U^\dag_\theta =
  H_{2d}(\bm\kappa,\Delta)\,,
\end{align}
where $\theta$ is a rotation angle about the $z$ axis, $U_\theta =
\cos\frac{\theta}{2}-i\sin\frac{\theta}{2}\sigma_z$ is a unitary
transformation acting on the spinors, and $R_\theta = \begin{pmatrix}\cos\theta &   \sin\theta \\ - \sin\theta & \cos\theta\end{pmatrix}$ is
rotation operation acting on $\bm\kappa$.  The rotational symmetry
determines that the linear
conductivity includes only two independent components, {\it i.e.}, the
diagonal component $\sigma^{(1);xx}_{2d}$ and the off-diagonal
component $\sigma^{(1);xy}_{2d}$. The other nonzero components can be found from 
\begin{align}
  \sigma^{(1);xx}_{2d}&=\sigma^{(1);yy}_{2d}\,, &
  \sigma^{(1);xy}_{2d}&=-\sigma^{(1);yx}_{2d}\,.
\end{align}
The off-diagonal components are nonzero because  the Berry
curvature at the Dirac point behaves as the vector potential of a magnetic monopole, and can
contribute to a Hall conductivity. 
For the third order conductivity, there are in all six independent
nonzero components, which can be taken to be $\sigma_{2d}^{(3);xxyy}$,
$\sigma_{2d}^{(3);xyxy}$, $\sigma_{2d}^{(3);xyyx}$,
$\sigma^{(3);yxyy}_{2d}$, $\sigma^{(3);yyxy}_{2d}$, and
$\sigma^{(3);yyyx}_{2d}$.  The other nonzero components are then given by
\begin{align}
  \sigma^{(3);xxxx}_{2d} &=\sigma^{(3);xxyy}_{2d}+\sigma^{(3);xyxy}_{2d}+\sigma^{(3);xyyx}_{2d}\,,\\
  \sigma^{(3);yxxx}_{2d} &=\sigma^{(3);yxyy}_{2d}+\sigma^{(3);yyxy}_{2d}+\sigma^{(3);yyyx}_{2d}\,,
\end{align}
and
\begin{align}
  \sigma^{(3);xxxx}_{2d}&=\sigma^{(3);yyyy}_{2d}\,, &\sigma^{(3);yxxx}_{2d}&=-\sigma^{(3);xyyy}_{2d}\,,\\
  \sigma^{(3);xxyy}_{2d}&=\sigma^{(3);yyxx}_{2d}\,, &\sigma^{(3);yxyy}_{2d}&=-\sigma^{(3);xyxx}_{2d}\,,\\
  \sigma^{(3);xyxy}_{2d}&=\sigma^{(3);yxyx}_{2d}\,, & \sigma^{(3);yyxy}_{2d}&=-\sigma^{(3);xxyx}_{2d}\,,\\
  \sigma^{(3);xyyx}_{2d}&=\sigma^{(3);yxxy}_{2d}\,, &  \sigma^{(3);yyyx}_{2d}&=-\sigma^{(3);xxxy}_{2d}\,.
\end{align}
For a single Dirac cone, the independent components $\sigma^{(1);xy}_{2d}$, $\sigma^{(3);yxyy}_{2d}$,
$\sigma^{(3);yyxy}_{2d}$, and $\sigma^{(3);yyyx}_{2d}$ are antisymmetric with respect to
$\{x\leftrightarrow y\}$, while the others, $\sigma^{(1);xx}_{2d}$, $\sigma_{2d}^{(3);xxyy}$,
$\sigma_{2d}^{(3);xyxy}$, and $\sigma_{2d}^{(3);xyyx}$ are
symmetric; we refer to these two different classes of tensor
components as ``antisymmetric'' and ``symmetric''
components, respectively.  Due to  inversion symmetry
\begin{align}
  \sigma_z H_{2d}(-\bm\kappa,\Delta)\sigma_z = H_{2d}(\bm\kappa,\Delta)\,,
\end{align}
and there is no second order response in the dipole approximation.

For 2D DFs, the sign of the mass parameter
determines the chirality, and the two different possibilities are connected through
\begin{align}
  U_mH_{2d}(R_m\bm\kappa,\Delta)U_m^\dag = H_{2d}(\bm\kappa,-\Delta)\,,
\end{align}
with $U_m = \frac{i}{\sqrt{2}}(\sigma_x-\sigma_y)$ and
$R_i=\begin{pmatrix} 0&-1 \\-1 &0\end{pmatrix}$. This
relation gives
$\sigma^{(n);dab\cdots}_{2d}(-\Delta)=\sigma^{(n);\bar d \bar a
  \bar b \cdots}_{2d}(\Delta)$ where the bar of a Roman letter means $\bar
d=y,x$ for $d=x,y$. Furthermore, utilizing the
consequences of rotational symmetry, we find that all
symmetric (antisymmetric) components are even (odd) functions of $\Delta$.

\subsection{Conductivities of gapped graphene}
We denote the conductivities that follow from the Hamiltonian $H_{gg;\tau}$ by $\sigma^{(1);da}_{gg;\tau}(\omega)$ and
$\sigma^{(3);dabc}_{gg;\tau}(\omega_1,\omega_2,\omega_3)$.
In the $\tau$ valley, the Hamiltonian connects to 
$H_{2d}(\bm\kappa,\Delta)$ through
\begin{align}
  H_{gg;\tau}(\bm\kappa,\Delta)=H_{2d}(R_\tau\bm \kappa,\Delta)\,, \label{eq:g2d}
\end{align}
with an orthogonal matrix
$R_\tau=\begin{pmatrix}0&\tau\\-1&0\end{pmatrix}$. From
Eq.~(\ref{eq:g2d}), the symmetric 
components satisfy $\sigma^{(n);da\cdots}_{gg;\tau}(\Delta)=\sigma_{2d}^{(n);\bar
  d\bar a\cdots}(\Delta)$, and antisymmetric components satisfy
$\sigma^{(n);da\cdots}_{gg;\tau}(\Delta)=\tau \sigma_{2d}^{(1);\bar
  d\bar a\cdots}(\Delta)$. Therefore, for gapped graphene only the
symmetric components survive, and they are
\begin{align}
  \sigma_{gg}^{(1);xx}(\omega) = 2 \sum_\tau
  \sigma^{(1);xx}_{gg;\tau}(\omega) = 4 \sigma^{(1);xx}_{2d}(\omega) \,,\label{eq:conn1}
\end{align}
where the prefactor $2$ comes from the spin degeneracy in gapped
graphene. Similarly the third order conductivities are
\begin{align}
  \sigma_{gg}^{(3);dabc}(\omega_1,\omega_2,\omega_3)
  &=
    4\sigma^{(3);dabc}_{2d}(\omega_1,\omega_2,\omega_3)\,, \label{eq:conn3}
\end{align}
for $dabc=xxyy$, $xyxy$, and $xyyx$.

The optical conductivities of gapped graphene under the linear
dispersion approximation have been studied, and analytical expressions
for them have been obtained \cite{APLPhotonics_4_034201_2019_Cheng}. For later use, we list the expressions in the clean
limit. The linear conductivity  is given by 
\begin{align}
  \sigma^{(1);xx}_{gg}(\omega) &=
  \frac{i\sigma_0}{\pi}\left[\frac{4E_c}{\hbar\omega} -
  \frac{4\Delta^2+(\hbar\omega)^2}{(\hbar\omega)^2}{\cal
      G}(E_c;\hbar\omega)\right]\,.\label{eq:cond1gg}
\end{align}
Here $\sigma_0=e^2/4\hbar$ is a universal conductivity,
$E_c=\text{max}\{|\Delta|,|\mu|\}$ is an effective gap parameter, and 
\begin{eqnarray}
  {\cal G}(E_c; \hbar\omega) &=& \ln\left|\frac{\hbar\omega+2E_c}{\hbar\omega-2E_c}\right| + i \pi \theta(|\hbar\omega|-2E_c)\,,
\end{eqnarray}
with $\theta(x)$ being the usual step function.  For the third order conductivity, the cyclic permutation symmetry on
$\{a\omega_1,b\omega_2,c\omega_3\}$ of
$\sigma^{(3);dabc}_{gg}(\omega_1,\omega_2,\omega_3)$ gives
\begin{equation}
  \sigma_{gg}^{(3);xxyy}(\omega_1,\omega_2,\omega_3) =\sigma^{(3);xyxy}_{gg}(\omega_2,\omega_1,\omega_3)=\sigma^{(3);xyyx}_{gg}(\omega_2,\omega_3,\omega_1)\,.
\end{equation}
The third order conductivity is then
\begin{eqnarray}
  (i\sigma_3)^{-1} \sigma^{(3);xxyy}_{gg}(\omega_1,\omega_2,\omega_3) 
  &=& F_1(\Delta;\hbar\omega_1,\hbar\omega_2,\hbar\omega_3){\cal
    G}(E_c;\hbar(\omega_1+\omega_2+\omega_3))\notag\\
  &+&
  F_2(\Delta;\hbar\omega_1,\hbar\omega_2,\hbar\omega_3) {\cal
    G}(E_c;\hbar(\omega_2+\omega_3) ) \notag\\
  &+&   F_3(\Delta;\hbar\omega_1,\hbar\omega_2,\hbar\omega_3) {\cal
    G}(E_c;\hbar(\omega_1+\omega_3) ) \notag\\
  &+& F_3(\Delta;\hbar\omega_1,\hbar\omega_3,\hbar\omega_2) {\cal
    G}(E_c;\hbar(\omega_1+\omega_2) ) \notag\\
  &+&  
  F_4(\Delta;\hbar\omega_1,\hbar\omega_2,\hbar\omega_3) {\cal
    G}(E_c;\hbar\omega_1) \notag\\
  &+& F_5(\Delta;\hbar\omega_1,\hbar\omega_2,\hbar\omega_3) {\cal
    G}(E_c;\hbar\omega_2) \notag\\
  &+& F_5(\Delta;\hbar\omega_1,\hbar\omega_3,\hbar\omega_2) {\cal
    G}(E_c;\hbar\omega_3)\,.\label{eq:cond3gg}
\end{eqnarray}
with $\sigma_3=\sigma_0(\hbar v_Fe)^2/\pi$. The coefficients $F_i$ are given by
\begin{equation}
  F_i(\Delta;\epsilon_1,\epsilon_2,\epsilon_3) = {\cal
    F}_{i0}(\epsilon_1,\epsilon_2,\epsilon_3) + \Delta^2 {\cal
    F}_{i2}(\epsilon_1,\epsilon_2,\epsilon_3) + \Delta^4 {\cal
    F}_{i4}(\epsilon_1,\epsilon_2,\epsilon_3)\,.
\end{equation}
All the expressions of ${\cal F}_{ij}$ are given in
Appendix~\ref{app:calF}. By setting $\Delta=0$ we get the third order
nonlinear conductivity for graphene as
\begin{align}
  \sigma^{(3);xxyy}_{gh}(\omega_1,\omega_2,\omega_3)
  &= \left.\sigma^{(3);xxyy}_{gg}(\omega_1,\omega_2,\omega_3)\right|_{\Delta=0}\,.\label{eq:sigma3gh}
\end{align}

We briefly discuss the asymptotic expression of these conductivities
as $\Delta\to\infty$. In that limit $E_c=\text{max}\{|\Delta|,|\mu|\}=\Delta$, and all involved
photon energies satisfy $\hbar\omega_i/E_c\to0$. As $\Delta\to\infty$, a
direct expansion in the small quantities $\hbar\omega_i/\Delta$ gives
\begin{align}
  \sigma^{(1);xx}_{gg}(\omega) &\to -i\sigma_0\frac{4 \hbar\omega}{3\pi
                            \Delta}\,,\label{eq:delta1}\\
  \sigma^{(3);xxyy}_{gg}(\omega_1,\omega_2,\omega_3)&\to -i\sigma_3 \frac{2\hbar(\omega_1+\omega_2+\omega_3)}{45\Delta^5}\,.\label{eq:delta3}
\end{align}

The effective gap parameters $E_c$ in Eq.~(\ref{eq:cond3gg}) appear only 
in functions of ${\cal G}$, which determine possible resonances
related to the interband  transitions. Considering the photon energies
involved in these functions, we note that the resonances can be associated with
one-photon, two-photon, and three-photon processes. Both the one-photon and
three-photon related resonances are similar to that of the linear
conductivity,  while the two-photon related resonance shows a different
behavior. Since $F_2(\Delta;\epsilon_1,\epsilon_2,\epsilon_3)=0$ for
$\epsilon_2+\epsilon_3=2\Delta$ and $F_3(\Delta;\epsilon_1,\epsilon_2,\epsilon_3)=0$ for
$\epsilon_1+\epsilon_3=2\Delta$, the two-photon related resonances disappear for an undoped system.  

\section{Conductivities for three-dimensional massless Dirac
  fermions\label{sec:sigmadf}}
With the symmetry properties of the conductivities for 2D massive DF in
one Dirac cone in hand, and with the analytic expressions of the
conductivities for 2D gapped graphene already determined, we
can now turn to the optical response of 3D massless DF. In this work, we
focus on the optical response of an isotropic 3D Dirac cone, although
more generally, of course, Dirac cones can be anisotropic; this is
briefly discussed in Appendix~\ref{app:sym}. For 3D massless DFs in a
single isotropic Dirac cone the Hamiltonian\cite{Opt.Express_27_38270_2019_Zhang} is
 \begin{align}
   H_{3d}(\bm k) = \hbar v_F \bm k\cdot\bm\sigma\,, \label{eq:h3d}
 \end{align}
where $\bm k=k_x\hat{\bm x}+k_y\hat{\bm y}+k_z\hat{\bm z}$ is a
 three dimensional wave vector. The
 two band energies are $\varepsilon_{\pm k}=\pm \hbar v_F|\bm k|$,
 which touch at $\bm k=\bm0$,  the Dirac point.

It is the conductivities following from this Hamiltonian in
Eq.~(\ref{eq:h3d}) that we study here, and we denote them by $\sigma^{(1);da}_{3d}(\omega)$ and
 $\sigma^{(3);dabc}_{3d}(\omega_1,\omega_2,\omega_3)$. The Hamiltonian
 $H_{3d}(\bm k)$ is spherical
 symmetric, and so the only independent nonzero component of the linear
 conductivity is  $\sigma^{(1);xx}_{3d}(\omega)$; for the third order conductivity, the
 independent nonzero components are the symmetric ones $\sigma^{(3);xxyy}_{3d}$,
 $\sigma^{(3);xyxy}_{3d}$, and $\sigma^{(3);xyyx}_{3d}$. All other components
 can be obtained either by
 \begin{align}
   \sigma^{(3);xxxx}_{3d}=\sigma^{(3);xxyy}_{3d}+\sigma^{(3);xyxy}_{3d}+\sigma^{(3);xyyx}_{3d}\,,
 \end{align}
 or  by permutation of the directions $\{x,y,z\}$.  Due to the cyclic permutation on
$\{a\omega_1,b\omega_2,c\omega_3\}$ of
$\sigma^{(3);dabc}_{3d}(\omega_1,\omega_2,\omega_3)$, and all nonzero
component can be written  in terms of
$\sigma^{(3);xxyy}_{3d}(\omega_1,\omega_2,\omega_3)$, which we
identify in the following.

The Hamiltonian for 3D massless DFs is connected to
that of 2D massive DFs through the relation $H_{3d}(\bm \kappa+{\Delta}/{(\hbar v_F)}\hat{\bm
  z})=H_{2d}(\bm \kappa,\Delta)$. In the calculation of both the
linear and nonlinear conductivities in the independent particle
approximation, the full response arises as the sum of the responses of
each independent particles, identified initially by its $\bm k$. Thus  the  response of 3D massless DFs to electric fields in the $x$ and $y$ directions
is equivalent to an ensemble of responses of 2D massive DFs with different gap parameters. In this manner the
linear conductivity can be written as
\begin{align}
  \sigma^{(1);xx}_{3d} &= \int
  \frac{dk_z}{2\pi}\sigma^{(1);xx}_{2d}(\hbar v_F k_z) =\frac{1}{\pi
  \hbar v_F}\int_0^\infty d\Delta
  \sigma^{(1);xx}_{2d}(\Delta)\notag\\
  &=  \frac{1}{4\pi \hbar v_F} \int_0^{\infty}d\Delta \sigma^{(1);xx}_{gg}(\Delta)\,,\label{eq:sigma12dto3d}
\end{align}
where we have used
$\sigma^{(1);xx}_{2d}(\Delta)=\sigma^{(1);xx}_{2d}(-\Delta)$ for
the second equal sign  and Eq.~(\ref{eq:conn1}) for the third equal sign. Similarly we have
\begin{align}
  \sigma^{(3);xxyy}_{3d} &=  \frac{1}{4\pi \hbar v_F} \int_0^{\infty}d\Delta \sigma^{(3);xxyy}_{gg}(\Delta)\,.\label{eq:sigma32dto3d}
\end{align}
Once these are determined, all other nonvanishing components of the
conductivities for 3D massless DFs follow from the symmetry properties
of those tensors.

Using the results for the conductivity of gapped graphene in
Eqs.~(\ref{eq:cond1gg}) and (\ref{eq:cond3gg}), the
integration can be done analytically, and the result is given in
Appendix~\ref{app:yn}. Because $\sigma^{(1);xx}_{gg}\propto
\Delta^{-1}$ in Eq.~(\ref{eq:delta1}), the integration in
Eq.~(\ref{eq:sigma12dto3d}) diverges; this is associated with the
assumption that the linear dispersion relation continues for all $\bm k$, no matter how large. Taking a cut-off energy
$E_A$ as the upper limit of the integration, to model the onset of more realistic band dispersion, the linear conductivity
of three dimensional Dirac fermions in one cone is
\begin{align}
  \sigma^{(1);xx}_{3d}(\omega)
  &= \sigma^{(1);xx}_{3d,reg}(\omega)-\frac{ie^2\hbar\omega}{12\pi^2\hbar^2v_F}\ln \frac{2E_A}{|\mu|}\,,\notag\\
  \sigma^{(1);xx}_{3d,reg}(\omega)
  &= \frac{ie^2}{24\pi\hbar^2  v_F}\frac{12 |\mu| ^2-5  (\hbar\omega)^2 + 3 (\hbar\omega)^2
    {\cal Z}(|\mu|;\hbar\omega)}{3\pi \hbar\omega}\,,\label{eq:cond1dm}
\end{align}
where  the function ${\cal Z}$ is given by
\begin{align}
  {\cal Z}(|\mu|;w) &=\ln |w^2-4\mu^2|-\ln \mu^2 -i \pi \text{sgn}(w)
                      \theta(|w|-2|\mu|)\notag\\
                    &={\cal T}\left(\frac{w}{|\mu|}\right)\,,
\end{align}
where 
\begin{align}
  {\cal T}(x) &=\ln \left|x^2-4\right| -i \pi \text{sgn}\left(x\right) \theta\left(x-2\right)\,,
\end{align}
with $\text{sgn}(x)$ the sign function. It is worth noting that $E_A$
is not an cut-off energy for the energies of the DFs, but
rather for the gap parameter; hence the expression in Eq.~(\ref{eq:cond1dm}) is not
exactly the same as those in literature that involve an energy cut-off 
\cite{Phys.Rev.B_93_235417_2016_Kotov,Opt.Commun._462_125319_2020_Ooi,Phys.Rev.B_100_85436_2019_Sonowal}. However,
 our result for the real part of the conductivity, which is the physically meaningful term, is consistent with 
 earlier results in literature. 

For the third order conductivity, the integration converges due to
$\sigma^{(3);xxyy}\propto \Delta^{-5}$ in Eq.~(\ref{eq:delta3}), and the conductivity of 3D Dirac fermions is
\begin{align}
  \sigma^{(3);xxyy}_{3d}(\omega_1,\omega_2,\omega_3)
  &= \frac{iv_F
    e^4}{16\pi^2} \Big\{ \frac{8}{45\hbar^3\omega_1\omega_2\omega_3}
    \notag\\ &+
     {\cal C}_{1}(\hbar\omega_1,\hbar\omega_2,\hbar\omega_3){\cal
    Z}(|\mu|;\hbar(\omega_1+\omega_2+\omega_3)) \notag\\
  &+ {\cal C}_{2}(\hbar\omega_1,\hbar\omega_2,\hbar\omega_3){\cal
     Z}(|\mu|;\hbar(\omega_2+\omega_3))\notag\\ & + {\cal C}_{3}(\hbar\omega_1,\hbar\omega_2,\hbar\omega_3){\cal
     Z}(|\mu|;\hbar(\omega_1+\omega_3))\notag\\  & + {\cal
      C}_{3}(\hbar\omega_1,\hbar\omega_3,\hbar\omega_2){\cal
     Z}(|\mu|;\hbar(\omega_1+\omega_2)) \notag\\ & +  
      {\cal C}_{4}(\hbar\omega_1,\hbar\omega_2,\hbar\omega_3) {\cal
     Z}(|\mu|;\hbar\omega_1)\notag\\
  &+ {\cal C}_{5}(\hbar\omega_1,\hbar\omega_2,\hbar\omega_3) {\cal
      Z}(|\mu|;\hbar\omega_2) \notag\\ &
 + {\cal C}_{5}(\hbar\omega_1,\hbar\omega_3,\hbar\omega_2) {\cal
      Z}(|\mu|;\hbar\omega_3)\Big\}\,,\label{eq:cond3dm}
\end{align}
where ${\cal C}_{i}$ is given by
\begin{align}
  {\cal C}_{i}(\hbar\omega_l, \hbar\omega_m, \hbar\omega_n){\cal Z}(|\mu|; \hbar\omega) &= \left[\sum_{j=0,2,4}{\cal F}_{ij}(\hbar\omega_l,\hbar\omega_m,\hbar\omega_n)\frac{(-\hbar\omega)^{j+1}}{2^{j+1}}\frac{1}{j+1}\right]{\cal Z}(|\mu|; \hbar\omega) \,.
\end{align}
Note that the coefficients ${\cal C}_i$ in $\sigma^{(3);xxyy}_{3d}$ satisfy 
\begin{align}
  &{\cal C}_{1}(\hbar\omega_1,\hbar\omega_2,\hbar\omega_3) + {\cal C}_{2}(\hbar\omega_1,\hbar\omega_2,\hbar\omega_3) + {\cal
  C}_{3}(\hbar\omega_1,\hbar\omega_2,\hbar\omega_3)
  + {\cal
      C}_{3}(\hbar\omega_1,\hbar\omega_3,\hbar\omega_2)\notag\\
  &+  
      {\cal C}_{4}(\hbar\omega_1,\hbar\omega_2,\hbar\omega_3) + {\cal C}_{5}(\hbar\omega_1,\hbar\omega_2,\hbar\omega_3) 
 + {\cal C}_{5}(\hbar\omega_1,\hbar\omega_3,\hbar\omega_2) =0\,.\label{eq:condc}
\end{align}

\section{Conductivities for different optical phenomena\label{sec:results}}
Since 3D massless DFs form an isotropic system, the  current density response can be written as
\begin{align}
  \bm J(t) &= \int \frac{d\omega}{2\pi} e^{-i\omega t}\sigma^{(1);xx}(\omega) \bm
             E_\omega \notag\\
           &+ \int\frac{d\omega_1d\omega_2d\omega_3}{(2\pi)^3}
             e^{-i(\omega_1+\omega_2+\omega_3)t}
             \left[\sigma^{(3);xxyy}_{3d}(\omega_1,\omega_2,\omega_3) \bm E_{\omega_1}
             (\bm E_{\omega_2}\cdot\bm E_{\omega_3})\right.\notag\\
           &+\left.\sigma^{(3);xyxy}_{3d}(\omega_1,\omega_2,\omega_3) \bm E_{\omega_2}
             (\bm E_{\omega_3}\cdot\bm E_{\omega_1})+\sigma^{(3);xyyx}_{3d}(\omega_1,\omega_2,\omega_3) \bm E_{\omega_3}
             (\bm E_{\omega_1}\cdot\bm E_{\omega_2})\right]\,,
\end{align}
where $\bm E_\omega = \int dt \bm E(t)e^{i\omega t}$ is the Fourier
transform of the electric field. In this section we consider the nature of this response for different optical phenomena.
\subsection{Several general properties of the conductivities}
We begin by discussing some general properties of the expressions for the
linear and third order conductivities in
Eqs.~(\ref{eq:cond1dm}) and (\ref{eq:cond3dm}).
\begin{enumerate}
\item For all the nonlinear phenomena we discuss,  the third order
  conductivity of 3D massless DFs exhibits features very similar to that of  graphene
   \cite{APLPhotonics_4_034201_2019_Cheng,NewJ.Phys._16_53014_2014_Cheng,*Corrigendum_NewJ.Phys._18_29501_2016_Cheng,Phys.Rev.B_91_235320_2015_Cheng},
   as we show below, including the appearance of resonances and divergences.   
   In 3D massless DFs the conductivities involve
   the function ${\cal Z}(|\mu|;\hbar\omega)$, instead of the function ${\cal
     G}(|\mu|;\hbar\omega)$ relevant for graphene. Both functions describe the interband
   optical transition, but they are weighted by different densities of states. However,
   there are always singularities at $|\hbar\omega|=2|\mu|$, around which the real
   part diverges logarithmically and the imaginary part shows a step
   function.  Similar to the frequency dependence of the nonlinear
   response of graphene, the third order conductivity of
   3D massless DFs involves photon energies
   $\hbar\omega_i$, $\hbar\omega_i+\hbar\omega_j$, and
   $\hbar(\omega_1+\omega_2+\omega_3)$, which appear in
   the second argument of the function ${\cal Z}(|\mu|;\hbar\omega)$. Thus, when
   any of these energies matches $2|\mu|$, a resonant interband transition may appear. When any
   of these energies is zero, an intraband divergence may appear and
   lead to a divergent conductivity value in the clean limit at zero temperature.

 \item Scaling all energies by the chemical potential, the third order conductivity can be written as 
   \begin{eqnarray}
     \sigma^{(3);dabc}_{3d}(|\mu|;\omega_1,\omega_2,\omega_3)
     &=&\frac{ v_F e^4}{16\pi^2|\mu|^3}S^{(3);dabc}_{3d}\left(\frac{\hbar\omega_1}{|\mu|},\frac{\hbar\omega_2}{|\mu|},\frac{\hbar\omega_3}{|\mu|}\right)\,,
   \end{eqnarray}
   where the dimensionless function $S_{3d}^{(3);dabc}$ can be obtained from
   $\sigma^{(3);dabc}_{3d}$. 
   To better understand the third order optical response of 3D
   massless DFs, we can compare it to that of graphene
   \cite{NewJ.Phys._16_53014_2014_Cheng,*Corrigendum_NewJ.Phys._18_29501_2016_Cheng}.
   If we introduce  an effective bulk conductivity of
   graphene by associating a thickness $d_{\text{eff}}\approx3.3$~{\AA} with a graphene
   sheet, that effective bulk third order conductivity $ \sigma^{(3);dabc}_{gh;eff}$ can be obtained
   from Eq.~(\ref{eq:sigma3gh}) by $ \sigma^{(3);dabc}_{gh;eff}=
   \sigma^{(3);dabc}_{gh}/d_{\text{eff}}$, and it can be written as
   \begin{align}
     \sigma^{(3);dabc}_{gh;eff}(|\mu|;\omega_1,\omega_2,\omega_3)
     &=\frac{\hbar v_F^2e^4}{4\pi d_{\text{eff}}|\mu|^4} S_{gh}^{(3);dabc}\left(\frac{\hbar\omega_1}{|\mu|},\frac{\hbar\omega_2}{|\mu|},\frac{\hbar\omega_3}{|\mu|}\right)\,,
   \end{align}
   where $S_{gh}^{(3);dabc}$ is a dimensionless function
   \cite{NewJ.Phys._16_53014_2014_Cheng,*Corrigendum_NewJ.Phys._18_29501_2016_Cheng}
   that can be obtained from $\sigma^{(3);dabc}_{gh}$. Besides the
   different detailed structures given in the
   dimensionless functions $S^{(3);dabc}_{3d}$ and $S^{(3);dabc}_{gh}$, the two conductivities above
   also show a different
   dependence on the Fermi velocity $v_F$ and the chemical potential $|\mu|$. Their ratio gives
   \begin{align}
     \frac{\sigma^{(3);xxyy}_{3d}}{\sigma^{(3);xxyy}_{gh;eff}} =
     \frac{d_{\text{eff}}|\mu|}{4\pi \hbar v_F} \frac{S^{(3);xxyy}_{3d}}{S_{gh}^{(3);xxyy}}\,.
   \end{align}
   The prefactor is inversely proportional to the Fermi velocity $v_F$
   and proportional to the chemical
   potential $|\mu|$. By taking the Fermi velocity to be that of
   graphene ($v_F=10^6$~m/s), the prefactor is
   about $0.04$ for $|\mu|=1$~eV. Therefore, the third optical
   conductivity of 3D massless DFs in one Dirac cone
   is about two orders of magnitude smaller than the corresponding
   effective bulk third order conductivity of graphene. Note that
   $\sigma^{(3);dabc}_{3d}$ is for one Dirac cone only; if there exists degeneracy $g$ of the Dirac cones, the third order
   conductivity $\sigma^{(3);dabc}_{3d}$ is $g$ times as large.

 \item When all involved frequencies satisfy 
   $\hbar\omega_i/|\mu| \ll 1$, the third order nonlinear response in a doped Dirac
   semimetal should be mostly due to the intraband transitions. This limit
   can be obtained by taking $\hbar\omega_i\to x \hbar\omega_i$ and
   $x\to0$, and we find an approximate conductivity is given by
   \begin{align}
     \sigma^{(3);xxyy}_{3d}(\omega_1, \omega_2, \omega_3)
     \approx  \frac{iv_Fe^4}{16\pi^2}
     \frac{8}{45\hbar^3\omega_1\omega_2\omega_3}\,.
   \end{align}
It is independent of the chemical potential $|\mu|$, showing a
different dependence on that quantity than that of graphene
($\propto|\mu|^{-1}$). Comparing this conductivity to the effective
bulk conductivity  of graphene, we find
\begin{align}
  \frac{\sigma^{(3);xxyy}_{3d}}{ \sigma^{(3);xxyy}_{gh;eff}}
  &= \frac{4|\mu|d_{\text{eff}}}{15\pi \hbar v_F}\,.
\end{align}
Taking the Fermi velocity to be that of graphene ($v_F=10^6$~m/s), for $|\mu|=1$~eV, the ratio is
about $0.042$. 

\item In the undoped limit as the chemical potential $\mu\to0$, the
  conductivities depend only on the frequencies. In this limit, the
  third order conductivity of graphene is very simple \cite{NewJ.Phys._16_53014_2014_Cheng,*Corrigendum_NewJ.Phys._18_29501_2016_Cheng}:
  $\sigma^{(3);xxyy}\propto
  [(\omega_1+\omega_2)(\omega_2+\omega_3)(\omega_3+\omega_1)(\omega_1+\omega_2+\omega_3)]^{-1}$.
  For 3D massless DFs, the expression for the third order
  conductivity in this limit is more 
  complicated. Although the function ${\cal Z}(|\mu|;\hbar\omega)$
  includes a term $\ln\mu^2$, it does not lead to any divergence because
  the term is cancelled out due to Eq.~(\ref{eq:condc}), thus the
  conductivity itself has no
  singularity at $|\mu|=0$, and is well behaved as
  $\mu\to0$.
  
\item Considering the dependence on the Fermi velocity $v_F$, the
  conductivities of graphene give $\sigma^{(n)}_{gh;eff}\propto
  v_F^{n-1}$, while those of 3D massless DFs give
  $\sigma^{(n)}_{3d}\propto v_F^{n-2}$. For graphene, the universal
  conductance appears in the linear optical response \cite{Science_320_1308_2008_Nair}. For Dirac
  fermions, the response independent  of the material parameter  should occur
  at second order, and in our simple model this is absent. But
  for Weyl semimetals, where  inversion symmetry is broken, the
  universal optical response does appear  in the circular photogalvanic effect \cite{Nat.Commun._8_15995_2017_Juan,Natl.Sci.Rev._6_206_2018_Moore}. 
\end{enumerate}

\subsection{Linear optical response}
For 3D massless DFs, the cutoff energy appears only in the imaginary
part of the linear conductivity. The real part in the clean limit is
given by
\begin{align}
  \text{Re}[\sigma^{(1);xx}_{3d}(\omega)] &= \frac{e^2\omega}{24\pi\hbar v_F}\theta(\hbar\omega-2|\mu|)\,,
\end{align}
which is proportional to the frequency $\omega$. This leads to a frequency
independent imaginary part of the susceptibility $\text{Im}[\chi(\omega)]=
\text{Re}[\sigma^{(1);xx}_{3d}(\omega)]/(\omega
\epsilon_0)={e^2}/(24\pi\hbar v_F\epsilon_0)$ for
$\hbar\omega>2|\mu|$, which is inversely proportional to the Fermi
velocity $v_F$.  Again taking the Fermi velocity to be the same as the
value for graphene, $v_F=10^6$~m/s,  the absorption coefficient is $\text{Im}[\chi(\omega)]\approx 0.36$. 

\begin{figure}[!htpb]
  \centering
  \includegraphics[width=5.cm]{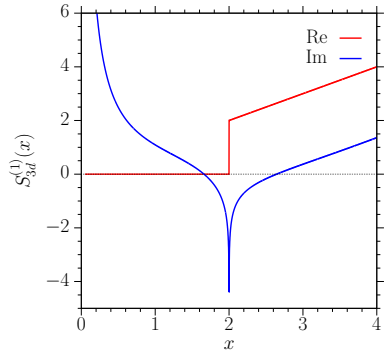}
  \caption{The $x$ dependence of $S_{3d}^{(1)}(x)$ in $0<x<4$. } 
  \label{fig:dmlinear}
\end{figure}

In the low frequency regime, the term involving the cutoff energy may contribute little due to its
prefactor $\hbar\omega$, and the main contribution comes from the Drude term 
\begin{align}
  \sigma^{(1);xx}_{3d}(\omega) \approx
  \frac{ie^2|\mu|^2}{6\pi^2\hbar^2 v_F}\frac{1}{\hbar\omega}\,.
\end{align}
It is proportional to the square of the chemical potential $|\mu|^2$, 
following the dependence of the density of states. The term
$\sigma^{(1);xx}_{3d,reg}(\omega)$ can be rewritten as
\begin{align}
  \sigma^{(1);xx}_{3d,reg}(\omega)
  &=    \frac{e^2|\mu|}{24\pi\hbar^2   v_F} S_{3d}^{(1)}\left(\frac{\hbar\omega}{|\mu|}\right)\,,
\end{align}
with a dimensionless function
\begin{align}
  S_{3d}^{(1)}(x) &=\frac{i}{\pi}\frac{ 12-5 x^2+3x^2 {\cal T}(x)}{3 x}\,.
\end{align}
Its real and imaginary parts are plotted in
Fig.~\ref{fig:dmlinear}. Around $x=2$,  there appears a
logarithmic divergence in its imaginary part and a step change in
its real part. For $x>2$ the real part is linearly dependent on $x$.

\subsection{Third harmonic generation}
The third order conductivity for THG satisfies
$\sigma^{(3);xxyy}_{3d}=\sigma^{(3);xyxy}_{3d}=\sigma^{(3);xyyx}_{3d}=\sigma^{(3);xxxx}_{3d}/3$. The
quantity $S_{3d}^{(3);xxyy}(x,x,x)$ is given by 
\begin{align}
  S_{3d}^{(3);xxyy}(x,x,x)&=\frac{2i}{135x^3}\left[12-5{\cal T}(x)+32{\cal
                       T}(2x)-27{\cal T}(3x)\right]\,.\label{eq:s3dthg}
\end{align}
Each ${\cal T}$ term is associated with one optical transition
involving photon energy $n\hbar\omega$ ($n=1$, $2$, or $3$). Similar to the
expression for the response tensor describing 
 THG in graphene, the prefactors of these terms
have different signs, indicating the existence of interference between these transitions. The real part is 
\begin{align}
  \text{Re}[S_{3d}^{(3);xxyy}(x,x,x)]
  &=\frac{2\pi \text{sgn}(x)}{135x^3}\left[-5 \theta(|x|-2) +
    32\theta(2|x|-2)-27\theta(3|x|-2)\right]\,.
\end{align}
For $x>2$, $ \text{Re}[S_{3d}^{(3)}(x,x,x)]=0$ gives a complete
cancellation due to interference. For graphene, the cancellation is
not complete \cite{NewJ.Phys._16_53014_2014_Cheng,Nat.Photon.___2018_Jiang}. 
\begin{figure}[!htpb]
  \centering
  \includegraphics[width=5.cm]{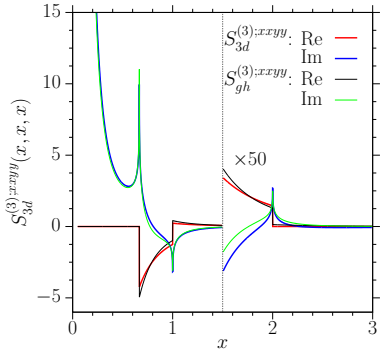}
  \caption{The $x$ dependence of $S_{3d}^{(3);xxyy }(x,x,x)$ and
    $S_{gh}^{(3);xxyy }(x,x,x)$. Values in the regime $x>1.5$ are scaled by $50$ times.} 
  \label{fig:dmthg}
\end{figure}
In Fig.~\ref{fig:dmthg}, we plot the spectra of
$S_{3d}^{(3);xxyy }$ and $S_{gh}^{(3);xxyy }$. They show very similar
amplitudes and structures.

We close the summary of our results by presenting the conductivity in the
limit of $\mu\to0$. It corresponds to taking $x\to\infty$ in
$S_{3d}^{(3);xxyy}(x,x,x)$; thus the real part is fully cancelled, and
the imaginary part is given by
\begin{align}
  \sigma^{(3);xxyy}_{3d}(\omega,\omega,\omega)|_{\mu=0} &= \frac{i
                                                 v_Fe^4 (6+32\ln
                                                 2-27\ln 3)}{540\pi^2 (\hbar\omega)^3}\,.
\end{align}
Finally we compare our results with those obtained in a velocity gauge
using Floquet states by Zhang {\it et al.}\cite{Opt.Express_27_38270_2019_Zhang} and Zhong {\it et al.}
\cite{Phys.B_555_81_2019_Zhong}. At
zero temperature, the real part of their results for one Dirac cone gives
\begin{align}
  \text{Re}[S_{3d}^{(3);xxyy}(x,x,x)]_{lit} = \frac{2\pi \text{sgn}(x)}{135x^3}\left[-4 \theta(|x|-2) +
  16\theta(2|x|-2)-27\theta(3|x|-2)\right]\,,
\end{align}
with the imaginary part obtained using Kramers-Kronig relations\cite{Opt.Express_27_38270_2019_Zhang}.
The results differ from ours in the first two factors for one and two
photon resonant processes, and the difference may arise from the
choice of  the velocity or length gauge to
describe the light-matter interaction. Considering the well-known problems that can result using the velocity gauge, a further
investigation is required to clarify what causes the different results
of these two methods.

\subsection{The Kerr effect and two photon absorption}
For a monochromatic laser beam, another important optical nonlinearity
invovles the corrections to the linear response due to the Kerr effect and two photon absorption, which are described by the tensor
$\sigma_{3d}^{(3);dabc}(-\omega,\omega,\omega)$. For the frequency set
$(-\omega,\omega,\omega)$, there are only two independent components
$\sigma_{3d}^{(3);xxyy}(-\omega,\omega,\omega)$ and
$\sigma_{3d}^{(3);xyyx}(-\omega,\omega,\omega)=\sigma_{3d}^{(3);xyxy}(-\omega,\omega,\omega)$.
Intraband divergences exist for this
third order conductivity, which are illustrated by
\begin{align}
 &  \begin{pmatrix}S^{(3);xyxy}_{3d}(-x,x+\delta_1,x+\delta_2) \\ 
   S^{(3);xxyy}_{3d}(-x,x+\delta_1,x+\delta_2)
   \end{pmatrix}
 =\frac{4\pi \text{sgn}(x)\theta(x^2-4)}{45w}B_d(x;\delta_1,\delta_2)
  + B_n(x) + \cdots\,.
\end{align}
Here the first term indicates all the intraband divergences with respect to
$\delta_1$ and $\delta_2$, but they are nonzero only when
one-photon absorption exists at $|x|>2$, which is consistent with the
general properties of intraband divergences
\cite{APLPhotonics_4_034201_2019_Cheng}.  The function $B_d$ is given by
\begin{align}
  B_d(x;\delta_1,\delta_2)
  &=\frac{ \begin{bmatrix}-3
    \\ 2\end{bmatrix}}{\delta_1\delta_2}
  +\frac{\begin{bmatrix}4x+3\delta_2\\-(x+2\delta_2)\end{bmatrix}}{\delta_1(x+\delta_2)(2x+\delta_2)}+\frac{\begin{bmatrix}9x+3\delta_1\\-(x+2\delta_1)\end{bmatrix}}{\delta_2(x+\delta_1)(2x+\delta_1)}\,.
\end{align}
The second term $B_n$ is well behaved and given by
\begin{align}
  B_n(x)&=\frac{1}{90x^3}\left(\begin{bmatrix}-31\\14\end{bmatrix}{\cal
  T}(-x)+ \begin{bmatrix}-65\\50\end{bmatrix}{\cal
  T}(x)+\begin{bmatrix}96\\-64\end{bmatrix}{\cal
  T}(2x)\right.\notag\\
  &\left.+x\begin{bmatrix}-52\\8\end{bmatrix}\frac{\partial {\cal
  T}(x)}{\partial x}+x^2\begin{bmatrix}-12\\8\end{bmatrix}\frac{\partial^2 {\cal
  T}(x)}{\partial x^2}-16\begin{bmatrix}1\\1\end{bmatrix}\right)\,.
\end{align}

\begin{figure}[!htpb]
  \centering
  \includegraphics[width=5.cm]{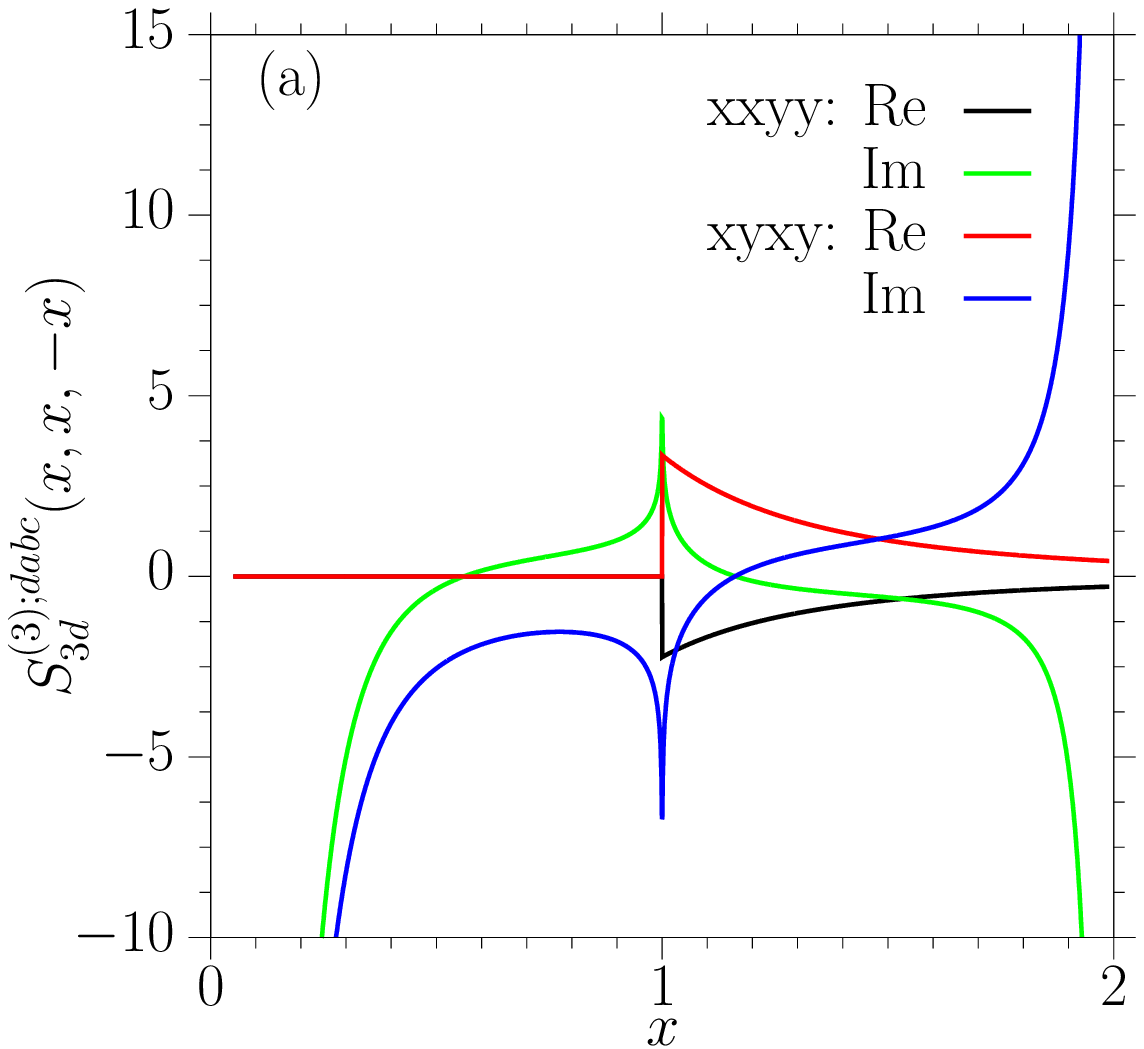}\includegraphics[width=5.cm]{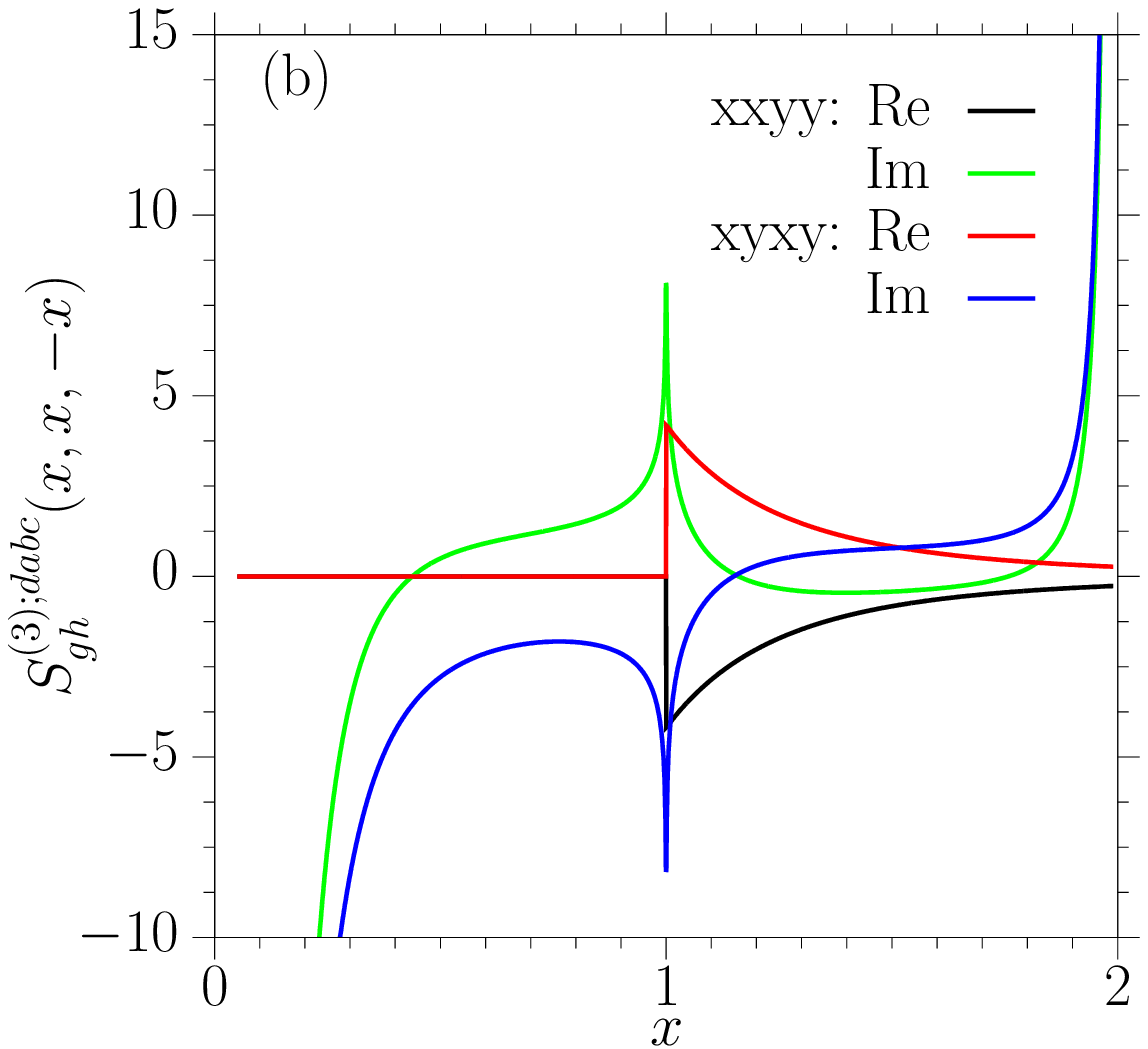}
  \caption{The $x$ dependence of (a) $S_{3d}^{(3);dabc}(x,x,-x)$ and
    (b) $S_{gh}^{(3);dabc}(x,x,-x)$ for the $xxyy$ and $xyxy$ components. } 
  \label{fig:dmkerr}
\end{figure}

In Fig.~\ref{fig:dmkerr} we plot  $S_{3d}^{(3);dabc}(-x,x,x)$ for
$0<x<2$, and compare it with
$S_{gh}^{(3);dabc}(-x,x,x)$. In general, both functions show very similar
structures and amplitudes, except for two obvious differences: (1) $\text{Im}[S_{3d}^{(3);xxyy}]$ diverges to $-\infty$
as $x\to2$, while $\text{Im}[S_{gh}^{(3);xxyy}]$ diverges to
$+\infty$; (2) For graphene the real parts of these two components satisfy
$\text{Re}[S_{gh}^{(3);xxyy}]=-\text{Re}[S_{gh}^{(3);xyxy}]$; however,
for 3D massless DFs  $S_{3d}^{(3);xxyy}$ this does not hold.
For $x>2$, the intraband divergences dominate, and in practice both the relaxation
processes and pulse shape effects will determine the magnitude of the
response. As a comparison, in the clean limit the results of Zhong
{\it et al.}\cite{Phys.B_555_81_2019_Zhong}, Zhang {\it et al.}
\cite{Opt.Express_27_38270_2019_Zhang}, and Ooi {\it et al.}
\cite{APLPhotonics_4_034402_2019_Ooi,Opt.Commun._462_125319_2020_Ooi} give
$\text{Re}[\sigma^{(3);xxxx}_{3d}]\propto\theta(\hbar\omega-2|\mu|)$, 
which contains no two photon absorption.\footnote{The results in
  Ref.~[\onlinecite{Phys.B_555_81_2019_Zhong}] have an obvious typo,
  as a comparison with those in
  Ref.~[\onlinecite{Opt.Express_27_38270_2019_Zhang}.]} We are not
sure whether or not such a difference occurs due to the different
choices for the light-matter interaction.

Next we present our results for two photon
carrier injection.  When one-photon absorption is absent ($x<2$), the two photon absorption
coefficient can be calculated through $\xi_2^{abcd}(\omega)
=3(\hbar\omega)^{-1}\text{Re}[\sigma^{(3);abcd}(-\omega,\omega,\omega)]$
\cite{NewJ.Phys._16_53014_2014_Cheng}. It can be written as
\begin{align}
  \begin{pmatrix}\xi_2^{xyxy}(\omega)\\\xi_2^{xxyy}(\omega)
  \end{pmatrix}
  &=\frac{v_F e^4}{240\pi |\mu|^4} \text{sgn}(\omega) X\left(\frac{\hbar\omega}{|\mu|};\frac{\hbar\delta_1}{|\mu|},\frac{\hbar\delta_2}{|\mu|}\right)\,,
\end{align}
with
\begin{align}
X(x;\delta_1,\delta_2) &= -\frac{4}{x^2} A_d(x;\delta_1,\delta_2) \theta(x^2-4)+\frac{1}{x^4}\begin{pmatrix}
  48\theta(x^2-1)-17\theta(x^2-4)\\-32\theta(x^2-1)+18\theta(x^2-4)
  \end{pmatrix}\,.
\end{align}
The first term comes from the intraband divergences, part of which
enters in the second term giving contributions proportional to $\theta(x^2-4)$. The first term exists only in the presence of
one-photon absorption ($x>2$), and physically the divergences are
induced by the stimulated Raman scattering process. For $1<x<2$ ({\it
  i.e.}, $|\mu|<\hbar\omega<2|\mu|$), two
photon absorption gives 
\begin{align}
  \begin{pmatrix}\xi_2^{xyxy}(\omega)\\\xi_2^{xxyy}(\omega)
  \end{pmatrix} = \frac{v_F e^4}{15\pi (\hbar\omega)^4} \begin{pmatrix}
    3\\-2
  \end{pmatrix}\,.
\end{align}
Compared to the results for graphene, the frequency dependence changes
from $\omega^{-5}$ to $\omega^{-4}$.

\subsection{Parametric frequency conversion}
\begin{figure}[!htpb]
  \centering
  \includegraphics[width=5.cm]{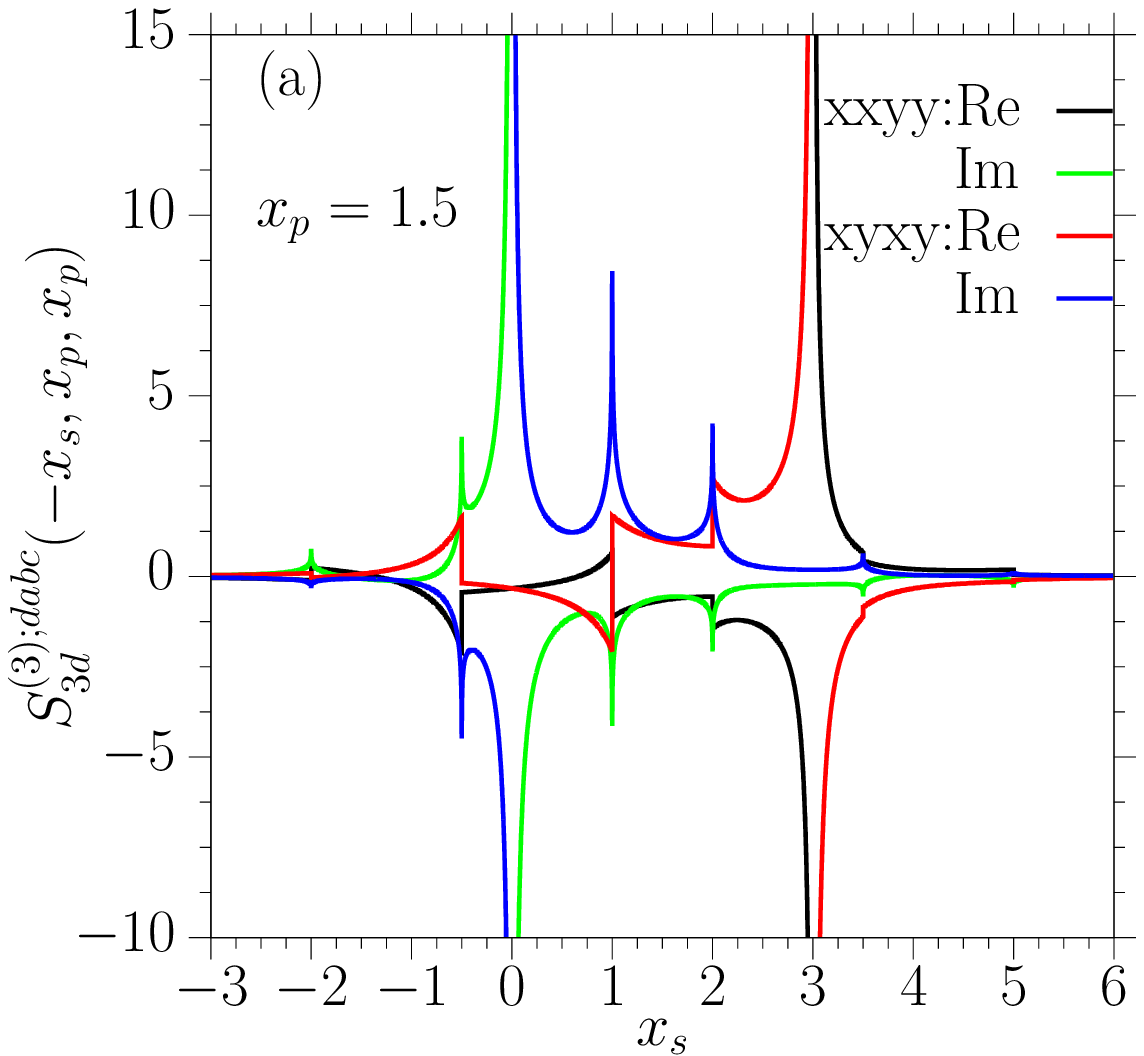}\includegraphics[width=5.cm]{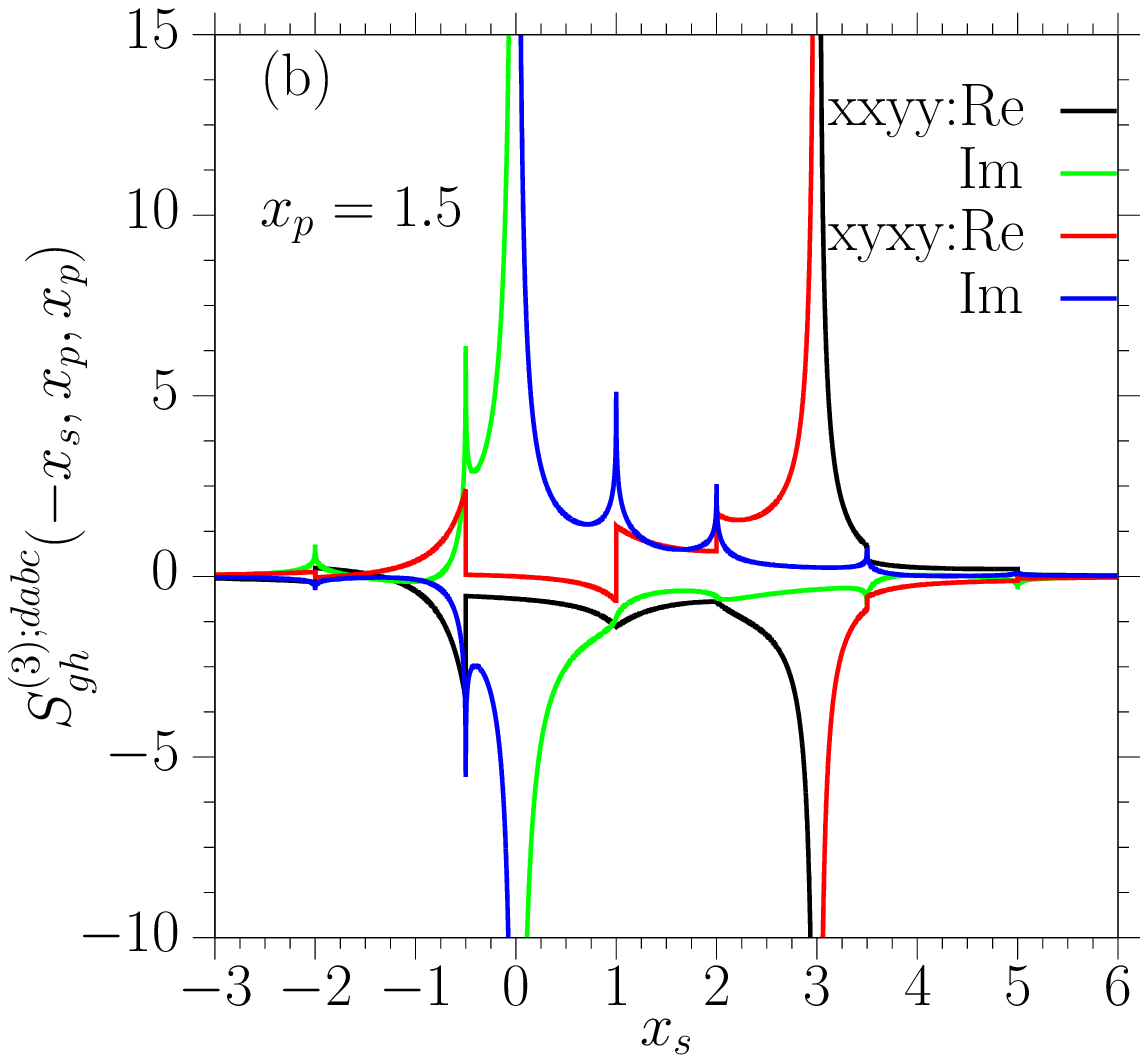}
  \caption{The $x_s$ dependence of the spectra for (a)
    $S_{3d}^{(3);dabc}(-x_s,x_p,x_p)$ for three dimension massless Dirac
    fermions and
    (b) $S_{gh}^{(3);dabc}(-x_s,x_p,x_p)$ for graphene. The pump
    frequency is chosen as $x_p=1.5$. } 
  \label{fig:dmfwm}
\end{figure}
When there are two laser beams, one with pump frequency $\omega_p$ and
the second with signal frequency $\omega_s$,  a new frequency $2\omega_p-\omega_s$ can
be generated through PFC; the current
density responsible for it is determined by
$\sigma^{(3);dabc}_{3d}(-\omega_s,\omega_p,\omega_p)$. For 3D massless
DFs, this process has only two independent components:
$\sigma^{(3);xxyy}_{3d}$ and 
$\sigma^{(3);xyxy}_{3d}=\sigma^{(3);xyyx}_{3d}$. Defining
$x_{s,p}\equiv\hbar\omega_{s,p}/\left|\mu\right|$, the term $S^{(3);dabc}_{3d}(-x_s,x_p,x_p)$ shows interband
divergences under the conditions $x_s=\pm 2$, $x_p=\pm 2$, $x_p=\pm 1$,
$x_p-x_s=\pm 2$,
or $2x_p-x_s=\pm 2$, and intraband divergences at $2x_p-x_s=0$, $x_s=0$. As an
illustration, we fix $x_p=1.5$ and show different components in
Fig.~\ref{fig:dmfwm}. The possible divergences appear
at $x_s=-2$, $-0.5$, $1$, $3$, and $3.5$ (interband), and at $x_s=0$ and $5$ (intraband). All these divergences exist for $S_{3d}^{(3);dabc}$, but
two of these divergences -- those at $x_s=1$
and $2$ -- are removed for $S_{gh}^{(3);dabc}$. Both conductivities  exhibit
similar amplitudes and structures. For the intraband divergences, that at $x_s=0$ is
associated with a field/current induced second harmonic generation,
and the other at $x_s=2x_p$ corresponds to two-color CCI, which is
discussed in the next section. Around these two divergences, the conductivities diverge as
$x_s^{-1}$ around $x_s\sim 0$, and $(x_s-2x_p)^{-1}$ as $x_s\sim 2x_p$. Obviously, the spectra diverge much faster
around intraband divergences than around interband divergences, where
the divergences are logarithmic.

\subsection{Two-color coherent current injection}
The intraband divergences of
$\sigma^{(3);dabc}_{3d}(-\omega,-\omega, 2\omega+\delta)$ as
$\delta\to0$ corresponds to a well known nonlinear phenomenon,
two-color coherent current injection, in which a quasi-static current
can be generated due to the interference of one-photon absorption and two-photon
absorption processes. The divergence means that the
current is continually injected, or
\begin{align}
  \frac{dJ^a(t)}{dt}
  &=  \eta^{abcd}_{3d}(\omega) E^b_{-\omega}E^c_{-\omega}E^d_{2\omega} + c.c.
\end{align}
with
\begin{align}
  \eta^{abcd}_{3d}(\omega) &=\lim_{\delta\to0}[-3i \delta 
                          \sigma^{(3);abcd}_{3d}(-\omega,-\omega,2\omega+\delta)]\,.
\end{align}
After simple algebra,  for $\omega>0$ we get
\begin{align}
  \begin{pmatrix} \eta_{3d}^{xxyy}(\omega) \\ \eta_{3d}^{xyyx}(\omega)
  \end{pmatrix} &=\frac{i v_Fe^4}{60\pi (\hbar\omega)^2}\left[\begin{pmatrix}-6\\4
                  \end{pmatrix}\theta(\hbar\omega-|\mu|)+\begin{pmatrix}2\\-3\end{pmatrix}\theta(\hbar\omega-2|\mu|)\right]\,.
\end{align}
The term involving $\theta(\hbar\omega-|\mu|)$ is associated with the
interference between the transition channels induced by a two-photon
absorption ($\omega + \omega$) and a one-photon absorption ($2\omega$), while
the other term involving $\theta(\hbar\omega-2|\mu|)$ is associated with the
interference of stimulated electronic Raman scattering (for photon
frequencies $2\omega$ and $-\omega$) and one-photon absorption
($\omega$). Compared to the injection in graphene, the injection
coefficients in 3D massless DFs are proportional to
$(\hbar\omega)^{-2}$, instead of $(\hbar\omega)^{-3}$ in graphene
\cite{NewJ.Phys._16_53014_2014_Cheng}; the relative amplitudes between
different components are also different.

\section{Conclusion and Discussion\label{sec:conclusion}}
We have calculated the linear and third order conductivities for a
single Dirac cone of 3D massless Dirac fermions. In our
simple model, we treat the light-matter interaction in the length gauge,
in which the kind of unphysical divergences associated with band
truncation that can appear in the velocity gauge do not arise. Analytic expressions for general input frequencies were
obtained in the clean limit at zero temperature. Utilizing these
expressions, we discussed in detail the frequency dependence of third
harmonic generation, the
Kerr effect and two photon absorption, parametric frequency
conversion, and two-color coherent current injection. The dimension affects
the optical response of Dirac fermions in several ways, and a comparison between two and three
dimensional massless Dirac fermions allows us to identify the
following qualitative features: (1) the  dependence on the Fermi
velocity $v_F$, which is the relevant material parameter in these systems, changes from $v_F^{n-1}$ in 2D to $v_F^{n-2}$ in 3D for the $n$th
order conductivity, (2)
the chemical potential dependence of the third order conductivity changes
from $\mu^{-1}$ to $\mu^0$ for a lightly doped sample, (3) the frequency
dependence of the two photon carrier injection changes from $\omega^{-5}$
to $\omega^{-4}$, (4) the frequency dependence of two color
current injection changes from $\omega^{-3}$ to $\omega^{-2}$, and (5) for
nonzero chemical potential, both frequency spectra show very similar
structures in general, but their amplitude can differ up to two order of
magnitude.

Although our
results are obtained in the clean limit at zero temperature, they
provide a general picture for third order response in three
dimensional massless Dirac fermions, and they can be treated as a
starting point for future study in nonlinear response of Dirac
 and Weyl semimetals. 

Finally, we discuss the inclusion of phenomenological relaxation parameters
and finite temperature, both of which are straightforward. For the third order conductivity of 
gapped graphene in our previous work \cite{APLPhotonics_4_034201_2019_Cheng}, the gap parameter appears in the
conductivities as functions of $1/E_c^{i}$ (i=1,3,5),
$\Delta^n{\cal G}(E_c;w)$, $\Delta^n{\cal H}(E_c;w) =
\frac{\partial}{\partial w}\left[\Delta^n {\cal G}(E_c;w)\right]$, and
$\Delta^n{\cal I}(E_c;w)=-\frac{\partial}{\partial w}\left[ \Delta^n{\cal
    H}(E_c;w)\right]$ for $n=0,2,4$. The integration of the latter two functions
with respect to $\Delta$ can be derived from those for $\Delta^n{\cal
  G}(E_c;w)$. The integrations of $\int_0^{E_A} E_c^{-i} d\Delta$ can also
be obtained easily. Therefore, the third order conductivity with
finite phenomenological relaxation parameters can be obtained by
replacing $\Delta^n{\cal
  G}(E_c;w)\to {\cal Y}_n(|\mu|;w)$, $\Delta^n{\cal H}(E_c;w) \to\frac{\partial}{\partial w} {\cal Y}_n(|\mu|;w)$, and
$\Delta^n{\cal I}(E_c;w)\to-\frac{\partial^2}{\partial w^2} {\cal
  Y}_n(|\mu|;w)$, and leaving the divergent terms with respect to $E_A$ in the integration
of $\int_0^{E_A} E_c^{-i} d\Delta$.  The complicated but analytic
expressions could be evaluated numerically. Starting from the
chemical potential dependence of the conductivity
$\sigma^{(1);xx}_{3d}(|\mu|;\omega)$ and
$\sigma^{(3);xxyy}_{3d}(|\mu|;\omega_1,\omega_2,\omega_3)$ at zero
temperature, the corresponding dependence at finite temperature can be constructed using the
technique presented earlier \cite{Phys.Rev.B_91_235320_2015_Cheng}. With this in hand, an
investigation of the effects of the relaxation parameter and finite
temperature on the optical conductivities of three dimensional Dirac
fermions can be undertaken.

However, we want to emphasize that even with such a treatment of phenomenological relaxation
parameters, and the consideration of finite temperature,  a detailed comparison with
experiments on materials exhibiting three dimensional massless Dirac fermions only makes sense for low light frequencies, due to small
energy range over which the assumption of a linear regime in the band
dispersion is valid. More generally, realistic calculations based on full band
structures will be required. Nonetheless, the study we have presented
here will serve as a benchmark for identifying when those full band
structure calculations show a significant difference from ideal Dirac
fermion behavior. 

\acknowledgements
This work has been supported by K.C.Wong Education
Foundation Grant No. GJTD-2018-08), Scientific research project of
the Chinese Academy of Sciences Grant No. QYZDB-SSW-SYS038, National
Natural Science Foundation of China Grant No. 11774340 and 61705227.
S.W.W. is supported by the National Key Research and Development
Program of China (Grant No 2019YFA0308404). J.E.S. is supported by the
Natural Sciences and Engineering Research Council of
Canada. J.L.C. acknowledges the support from ``Xu Guang'' Talent
Program of CIOMP.

\appendix

\section{Comparing responses\label{app:sym}}
We consider the relation between the optical conductivities of two
different systems with Hamiltonians, $H^{A}(\bm k)$ and
$H^{B}(\bm k)$, that are connected via a unitary matrix $U$ and a
real matrix $R$ through
\begin{align}
  UH^{A}(R\bm k)U^\dag = H^{B}(\bm k)\,. \label{eq:consym}
\end{align}
The dynamics of these two systems can be described by density matrices
$\rho_{\bm k}^A(t)$ and $\rho_{\bm k}^B(t)$. Under the application of
electric field $\bm E(t)$, they satisfy the equation of
motion \cite{NewJ.Phys._16_53014_2014_Cheng,*Corrigendum_NewJ.Phys._18_29501_2016_Cheng}
\begin{align}
  \hbar\partial_{t} \rho_{\bm k}^A(t)
  &= -i [  H^A(\bm k), \rho^A_{\bm k}(t) ] + e\bm
    E(t)\cdot\bm\nabla_{\bm k}\rho^A_{\bm k}(t)\,,\label{eq:sbeA}\\
  \hbar\partial_{t} \rho_{\bm k}^B(t)
  &= -i [  H^B(\bm k), \rho^B_{\bm k}(t) ] + e\bm
    E(t)\cdot\bm\nabla_{\bm k}\rho^B_{\bm k}(t)\,.\label{eq:sbeB}
\end{align}
To clearly indicate the field that leads to the response, we denote the
solutions of these two equations as $\rho^{A/B}_{\bm k}(t;\bm E(t))$. 
The current density responses are  functionals of the field $\bm
E(t)$, and can be calculated as
\begin{align}
  \bm J^A(t; \bm E(t))&= -\frac{e}{\hbar}\sum_{\bm
                          k}\text{Tr}\left[\rho_{\bm k}^A(t;\bm E(t))\bm\nabla_{\bm k}H^A(\bm   k)\right]\,.\label{eq:jA}\\
  \bm J^B(t; \bm E(t))&= -\frac{e}{\hbar}\sum_{\bm
                          k}\text{Tr}\left[\rho_{\bm k}^B(t; \bm E(t))\bm\nabla_{\bm k}H^B(\bm
            k)\right]\,.\label{eq:jB}
\end{align}

Now we determine the connection between $\rho^A_{\bm k}(t; \bm E(t))$ and
$\rho^B_{\bm k}(t; \bm E(t))$ induced by the relation in Eq.~(\ref{eq:consym}). Considering a transformation
\begin{align}
  \overline{\rho}_{\bm  k}(t)=U\rho^A_{R\bm k}(t; \bm E(t))U^{-1}\,,\label{eq:x}
\end{align}
from Eq.~(\ref{eq:sbeA}), the dynamics of
$\overline{\rho}_{\bm k}(t)$ is
\begin{align}
  \hbar\partial_{t} \overline{\rho}_{\bm k}(t)
  &= -i [UH^A(R\bm k)U^{-1}, \overline{\rho}_{\bm k}(t) ] + e[R\bm
    E(t)]\cdot\bm\nabla_{\bm k}\overline{\rho}_{\bm k}(t)\,.\label{eq:sbebar}
\end{align}
Utilizing Eq.~(\ref{eq:consym}) it is transformed into
Eq.~(\ref{eq:sbeB}), and we can find the solution is 
\begin{align}
\overline{\rho}_{\bm k}(t) =  \rho^B_{\bm k}(t; R\bm E(t))\,,
\end{align}
Then from Eq.~(\ref{eq:x}) the connection between $\rho^A_{\bm k}(t)$ and $\rho^B_{\bm k}(t)$ is
\begin{align}
  \rho^B_{\bm k}(t; R\bm E(t)) &=  U\rho^A_{R\bm k}(t; \bm E(t))U^{-1} \,.\label{eq:y}
\end{align}
In Eq.~(\ref{eq:jB}) by replacing $\bm E(t) \to R\bm E(t)$ and utilizing
Eq.~(\ref{eq:y}) and then comparing to Eq.~(\ref{eq:jA}), we get
\begin{align}
  \bm J^A(t;\bm E(t)) = |R|\left(R^T\right)^{-1}\bm J^B(t;R\bm E(t))\,.
\end{align}  
Note that for all of this analysis $R$ is not limited to be a
orthogonal matrix, and therefore such transformation can be used to
connect the response of an
anisotropic Dirac cone, {\it i.e.} $H(\bm k)=\hbar v_f \bm k\cdot
R\cdot\bm\sigma$,  to that of an isotropic cone $H(\bm k)=\hbar v_f
\bm k\cdot\bm \sigma$. 

For a weak electric field $\bm E(t)$, the induced current density can be expanded in a
power series of this field, and the expansion coefficients are the conductivity tensors. As an example, if the  matrix $R$ corresponds to an
orthogonal matrix $R^T=R^{-1}$, the linear conductivity and third order
conductivity of these two systems satisfy 
\begin{align}
 \sigma_A^{(1);da} &= R^{d d^\prime} R^{a a^\prime} \sigma_B^{(1);d^\prime a^\prime}  \,,\\
\sigma_A^{(3);dabc} &= R^{d d^\prime} R^{a a^\prime}R^{bb^\prime}R^{cc^\prime}
  \sigma_B^{(3);d^\prime a^\prime b^\prime c^\prime}\,.
\end{align}

\section{Expressions of ${\cal F}_{ij}$\label{app:calF} for gapped graphene}
Using $\epsilon_{ij}=\epsilon_i+\epsilon_j$ and $\epsilon=\epsilon_1+\epsilon_2+\epsilon_3$, we write
\begin{align}
  {\cal F}_{ij}(\epsilon_1,\epsilon_2,\epsilon_3)
  &=\frac{\overline{\cal  F}_{ij}(\epsilon_1,\epsilon_2,\epsilon_3)
    }{6\epsilon_1^2\epsilon_2^2\epsilon_3^2\epsilon_{12}\epsilon_{23}\epsilon_{31}\epsilon}\,,
\end{align}
where $\overline{\cal  F}_{ij}(\epsilon_1,\epsilon_2,\epsilon_3)$ are
given by
\begin{align}
\overline {\cal F}_{10}(\epsilon_1,
  \epsilon_2,\epsilon_3)&=\epsilon^2\left[3\epsilon_1^3\epsilon_{23}+(-\epsilon_2\epsilon_3+2\epsilon_1^2-\epsilon_1\epsilon_{23})\epsilon_{23}^2+\epsilon_1\epsilon_2\epsilon_3(2\epsilon_{23}-\epsilon_1)\right]\,,\\
  \overline{\cal F}_{12}(\epsilon_1,\epsilon_2,\epsilon_3) &=
  -8\left[3\epsilon_1^2\epsilon_2\epsilon_3+\epsilon_1^3\epsilon_{23}-\epsilon_{23}^2(\epsilon_2\epsilon_3+\epsilon_1\epsilon_{23})\right]\,,\\
  \overline{\cal F}_{14}(\epsilon_1,\epsilon_2,\epsilon_3) &=-16(\epsilon_2\epsilon_3+\epsilon_1\epsilon_{23})\,.
\end{align}
\begin{align}
  \overline{\cal F}_{20}(\epsilon_1,
  \epsilon_2,\epsilon_3) &=
  \epsilon_{12}\epsilon_{13}\epsilon_{23}^4\,,\\
  \overline{\cal F}_{22}(\epsilon_1,
  \epsilon_2,\epsilon_3)& =-8
  \epsilon_{12}\epsilon_{13}\epsilon_{23}^2 \,, \\
  \overline{\cal F}_{24}(\epsilon_1,
  \epsilon_2,\epsilon_3) &=16
  \epsilon_{12}\epsilon_{13}\,.
\end{align}
\begin{align}
\overline{\cal F}_{30}(\epsilon_1,
  \epsilon_2,\epsilon_3) &=-
  \epsilon_{12}\epsilon_{13}^2\epsilon_{23}\left[3\epsilon_1^2+2\epsilon_1\epsilon_2-3\epsilon_2\epsilon_3+(2\epsilon_1-\epsilon_3)\epsilon_{23}\right]\,,\\
  \overline{\cal F}_{32}(\epsilon_1,
  \epsilon_2,\epsilon_3)& =8
  \epsilon_{12}(\epsilon_1-\epsilon_3)\epsilon_{23}(\epsilon+\epsilon_2) \,, \\
  \overline{\cal F}_{34}(\epsilon_1,
  \epsilon_2,\epsilon_3) &= 16\epsilon_{12}\epsilon_{23}\,.
\end{align}
\begin{align}
\overline  {\cal F}_{40}(\epsilon_1,
  \epsilon_2,\epsilon_3)
  &=\epsilon_1^2\left[\epsilon_2\epsilon_3(\epsilon+\epsilon_{23})^2+\epsilon(\epsilon-\epsilon_{23})\epsilon_{23}(3\epsilon+\epsilon_{23})\right]
  \,,\\
  \overline{\cal F}_{42}(\epsilon_1,
    \epsilon_2,\epsilon_3) &=-8\left[\epsilon^3\epsilon_{23}-\epsilon\epsilon_{23}^3+\epsilon_2\epsilon_3(-3\epsilon^2+\epsilon_{23}^2)\right]\,,\\
  \overline{\cal F}_{44}(\epsilon_1,
  \epsilon_2,\epsilon_3)&=-16(-\epsilon_2\epsilon_3+\epsilon_{23}\epsilon)\,.
\end{align}
\begin{align}
\overline  {\cal F}_{50}(\epsilon_1,\epsilon_2,\epsilon_3) &=
  -\epsilon_2^2\left[\epsilon\epsilon_1(\epsilon_{23}+\epsilon_3)^2+\epsilon_2\epsilon_3\epsilon_{23}(\epsilon_{23}+3\epsilon_3)\right]\,,\\
\overline  {\cal F}_{52}(\epsilon_1,\epsilon_2,\epsilon_3)
  &=8\left(-\epsilon_3^3\epsilon_{23}+\epsilon_3\epsilon_{23}^3-3\epsilon_1\epsilon_3^2\epsilon+\epsilon_1\epsilon_{23}^2\epsilon\right)\,,\\
\overline  {\cal F}_{54}(\epsilon_1,\epsilon_2,\epsilon_3)
  &=-16(\epsilon_3\epsilon_{23}+\epsilon_1\epsilon)\,.
\end{align}

\section{Conductivity for 3D Dirac
  Fermions\label{app:yn}}
The linear conductivity and third order conductivity of three dimensional Dirac fermions are constructed from
Eqs.~(\ref{eq:sigma12dto3d}) and (\ref{eq:sigma32dto3d}),
respectively. The upper limit of the integration is
infinity, and thus it is necessary to introduce a cutoff energy $A$ to analyse the integration
\begin{align}
  I^{(n)}(A) = \frac{1}{4\pi \hbar v_F}\int_0^Ad\Delta \sigma^{(n)}_{gg}(\Delta)\,,
\end{align}
and then $\sigma^{(n)}_{3d}=\lim\limits_{A\to\infty}I^{(n)}(A)$. As
$\Delta\to\infty$, From Eqs.~(\ref{eq:delta1}) and (\ref{eq:delta3}) we have
$\sigma^{(1);xx}_{gg}(\Delta\to\infty)
\to - {4 i\sigma_0 \hbar\omega}/(3\pi)\Delta^{-1}$
and
$\sigma^{(3);xxyy}_{gg}(\Delta\to\infty)
\sim \Delta^{-5}$. 
It is obvious that $I^{(1);xx}(A\to\infty)$ diverges as $\ln A$ and
$I^{(3);xxyy}(A\to\infty)$ converges.

The $\Delta$ dependence in the conductivities of gapped graphene appears
in $\Delta$ or $\Delta^n{\cal G}(E_c;w)$ for $n=0,2,4$. By extending the
definition of ${\cal G}(E_c;w)$ to a complex $w=w_r+iw_i$ 
we get
\begin{align}
  {\cal G}(E_c;w) = i\pi + {\cal L}(E_c; w) - {\cal L}(-E_c; w)\,
\end{align}
with
\begin{align}
  {\cal L}(x;w_r+iw_i) &=\frac{1}{2}\ln \left[\left(w_r+2x\right)^2+w_i^2\right]- i
                         \arctan\frac{w_r+2x}{w_i}\,,
\end{align}
As $w_i\to0^+$, it becomes
\begin{align}
  {\cal L}(x;w_r) &=\ln |w_r+2x|- i \frac{\pi}{2} \text{sgn}(w_r+2x)\,.
\end{align}
with $\text{sgn}(x)$ the sign function.  

For the term $\Delta^n {\cal
  G}(E_c;w)$,  the integration is
\begin{align}
  \int_0^{E_A} x^n {\cal G}(\text{max}\{|\mu|,x\}; w) dx
  &= \int_0^{|\mu|}  x^n {\cal G}(|\mu|;w) dx + \int_{|\mu|}^{E_A} x^n {\cal G}(x; w) dx
    \notag\\
  &= \frac{|\mu|^{n+1}}{n+1}{\cal G}(|\mu|; w)+
    {\cal K}_n(E_A; w) -{\cal K}_n(|\mu|;w)\,,
\end{align}
with
\begin{align}
  {\cal K}_n(x;w) &= \frac{1}{n+1}\left[x^{n+1}{\cal G}(x;w) - {\cal
                    Q}_n(x; w)\right]\,,\\
  {\cal Q}_n(x;w) &=  \frac{(-w)^{n+1}}{2^{n+1}} \left[{\cal
                    L}(x;w)+(-1)^n{\cal L}(-x;w)\right] \notag\\
                  &+\frac{1}{2^{n+1}}
                    \sum_{m=1}^{n+1}C_{n+1}^m \frac{(-w)^{n+1-m}}{m}\left[(w+2x)^m-(-1)^{n+1}(w-2x)^m\right]\,.
\end{align}
Taking $A\to\infty$, ${\cal K}_n(A;w)$ diverges as $\propto \ln
(2E_A)$, $E_A^2$, and $E_A^4$ for $n=0,2,4$. We collect all divergent terms
into ${\cal R}_n(A;w)$ and write ${\cal K}_n(E_A;w)=\overline{\cal
  K}_n(w) + {\cal R}_n(E_A;w)$ with
\begin{align}
  \overline{\cal K}_0(w) &= 0\,,\quad \overline{\cal K}_2(w)
  =-\frac{1}{8}w^3\,,\quad \overline{\cal K}_4(w) =-\frac{5}{192}w^5\,.
\end{align}
Therefore the integration becomes
\begin{align}
  \int_0^{E_A} x^n {\cal G}(\text{max}\{|\mu|,x\}; w) dx 
  &= {\cal Y}_n(|\mu|;w) + {\cal R}_n(A;w)\,, 
\end{align}
with
\begin{align}
{\cal Y}_n(|\mu|;w) &=\overline{\cal K}_n(w) + \frac{1}{n+1}{\cal  Q}_n(|\mu|;w)
\end{align}

Now we can construct the conductivity $\sigma^{(n)}_{3d}$ from that of $\sigma^{(n)}_{gg}$ by replacing $\Delta^n {\cal G}(E_c;w)$ with
${\cal Y}_n(|\mu|;w)$. For the linear conductivity
$\sigma^{(1);xx}_{3d}(\omega)$, the divergent term can be obtained
from Eq.~(\ref{eq:delta1}) directly. 
Based on Eq.~(\ref{eq:cond3gg}), in the clean limit the third order conductivity for Dirac fermions is
\begin{eqnarray}
&&  \sigma^{(3);xxyy}_{3d}(|\mu|;\omega_1,\omega_2,\omega_3)\notag\\
  &=& \frac{iv_F
    e^4}{16\pi^2}\sum_{j=0,2,4}\Big\{ {\cal F}_{1j}(\hbar\omega_1,\hbar\omega_2,\hbar\omega_3){\cal
    Y}_j(|\mu|;\hbar(\omega_1+\omega_2+\omega_3)) \notag\\
  &&+ {\cal F}_{2j}(\hbar\omega_1,\hbar\omega_2,\hbar\omega_3){\cal
     Y}_j(|\mu|;\hbar(\omega_2+\omega_3)) + {\cal F}_{3j}(\hbar\omega_1,\hbar\omega_2,\hbar\omega_3){\cal
     Y}_{j}(|\mu|;\hbar(\omega_1+\omega_3))\notag\\  && + {\cal
      F}_{3j}(\hbar\omega_1,\hbar\omega_3,\hbar\omega_2){\cal
     Y}_j(|\mu|;\hbar(\omega_1+\omega_2)) +  
      {\cal F}_{4j}(\hbar\omega_1,\hbar\omega_2,\hbar\omega_3) {\cal
     Y}_j(|\mu|;\hbar\omega_1)\notag\\
  &&+ {\cal F}_{5j}(\hbar\omega_1,\hbar\omega_2,\hbar\omega_3) {\cal
      Y}_j(|\mu|;\hbar\omega_2) 
 + {\cal F}_{5j}(\hbar\omega_1,\hbar\omega_3,\hbar\omega_2) {\cal
      Y}_j(|\mu|;\hbar\omega_3)\Big\}\,.
\end{eqnarray}
It can be simplified in terms of the function ${\cal L}(x; w)+{\cal
  L}(-x; w)$, and we then get the expression in Eq.~(\ref{eq:cond3dm}).

%

\end{document}